\documentclass{article}
\usepackage[utf8]{inputenc}
\usepackage{authblk}
\usepackage{setspace}
\usepackage[margin=1.25in]{geometry}
\usepackage{graphicx}
\graphicspath{ {./figures/} }
\usepackage{subcaption}
\usepackage{amsmath}

\usepackage[colorlinks=true, allcolors=blue]{hyperref}

\usepackage[style=ieee, 
citestyle=numeric-comp,
sorting=none]{biblatex}
\title{The strength of structural diversity in online social networks}

\author[1,2]{Yafei Zhang}
\author[1]{Lin Wang}
\author[2*]{Jonathan J. H. Zhu}
\author[1,3*]{Xiaofan Wang}
\author[4]{Alex {\lq Sandy\rq} Pentland}

\affil[1]{Department of Automation, Shanghai Jiao Tong University, and Key Laboratory of System Control and Information Processing, Ministry of Education of China, Shanghai 200240, China}
\affil[2]{Department of Media and Communication, and School of Data Science, City University of Hong Kong, Hong Kong S.A.R., China}
\affil[3]{Department of Automation, Shanghai University, Shanghai 200444, China}
\affil[4]{Media Lab, Massachusetts Institute of Technology, Cambridge, MA 02139, USA}

\date{}

\begin{document}

\maketitle


\begin{abstract}
Understanding the way individuals are interconnected in social networks is of prime significance to predict their collective outcomes.
Leveraging a large-scale dataset from a knowledge-sharing website, this paper 
presents an exploratory investigation of the way to depict structural diversity in directed networks and how it can be utilized to predict one's online social reputation.
To capture the structural diversity of an individual, we first consider the number of weakly and strongly connected components in one's contact neighborhood and further take the coexposure network of social neighbors into consideration.
We show empirical evidence that the structural diversity of an individual is able to provide valuable insights to predict personal online social reputation, and the inclusion of coexposure network provides an additional ingredient to achieve that goal.
After synthetically controlling several possible confounding factors through matching experiments, structural diversity still plays a nonnegligible role in the prediction of personal online social reputation.
Our work constitutes one of the first attempts to empirically study structural diversity in directed networks and has practical implications for a range of domains, such as social influence and collective intelligence studies.
\end{abstract}


\section{Introduction}
Recent years have witnessed the emergence and rapid proliferation of many social applications and media platforms. 
As the backbone of so many online social systems, network structure is becoming a complex and subtle force that drives the dynamics of a wide variety of social processes.
In some cases, we seek to leverage social networks to maintain social capital, encourage the adoption of new products or promote positive behaviors like cooperation and physical exercise \cite{ellison2007benefits, ugander2012structural, banerjee2013diffusion, mani2013inducing, centola2010spread, aral2017exercise, althoff2017online, proestakis2018network, steinert2015online, jin2019emergence}, while in others to eliminate the spread of infectious diseases and fake news or change negative behaviors like conflict and unhealthy eating \cite{christakis2007spread, madan2010social, pastor2015epidemic, paluck2016changing, vosoughi2018spread}. 

A wealth of studies suggests that the socio-economic characteristics of individuals or communities are closely related to their network locations \cite{lin22building, wasko2005should, eagle2010network, luo2017inferring, leo2016socioeconomic, bollen2017happiness, blumenstock2015predicting, gao2019computational}.
As evidenced in the literature, connected individuals generally show similar patterns in terms of diverse social interests and activities \cite{mcpherson2001birds, aral2009distinguishing, centola2011experimental, christakis2013social, liang2018birds}.
However, the information redundancy and prevalence of similarity in one's social realm may limit his/her potential of exposure to diverse information and interaction with people from different backgrounds, thereby reducing the efficiency of social networks, preventing the diffusion of innovative ideas, weakening the power of social influence
and undermining the wisdom of crowds \cite{granovetter1973strength, woolley2010evidence, muchnik2013social, park2018strength, moore2021inclusivity}.

The recent availability of vast and fine-grained data of human activities in online social networks provides an unprecedented opportunity to investigate the nuanced or subtle social effects induced by social context diversity. 
Structural diversity, which aims to quantify the diversity of one's social context from the view of component count among social neighbors, has been shown useful in predicting specific social processes in undirected networks \cite{ugander2012structural, aral2017exercise}.
While for directed networks, the strength of structural diversity still remains an open question \cite{su2020experimental}.
In directed networks, the network connectivity patterns become more complicated, and network ties may function quite differently according to the direction \cite{liang2018birds, almaatouq2016you}.
Moreover, exposure to similar users could have a better chance to result in the similarity between two individuals even without an explicit social tie between them \cite{dong2018social}, which is largely overlooked by previous studies on structural diversity.

Online social reputation is the consensus public opinion of an individual or entity based on the ratings from members in a social network \cite{liu2017identifying, josang2007survey, gao2015group}.
It helps to foster trust among online users and is one of the most valuable assets in our online social lives \cite{gao2017evaluating, resnick2000reputation}.
Using data from an online knowledge-sharing platform, this paper addresses the problem of how to quantify the structural diversity of an individual in directed networks and the role of it in empirically predicting personal social reputation.
We first consider the weak and strong connectivity patterns among one’s social contacts and find that personal social reputation is positively correlated with the number of weakly or strongly connected components in one’s contact neighborhood.
Conditional on how many followers one has, individuals whose followers come from more diverse social backgrounds tend to have higher social reputations on the platform.
Regression analysis indicates that structural diversity measures via weak and strong connectivity among one's neighbors yield better predictions of personal social reputation than the number of followers one has.
We further take the coexposure network of one's social neighbors into consideration, which goes beyond the sheer number of social contacts or connected social components in one's contact neighborhood and provides an additional ingredient to predict personal social reputation.

To eliminate the effects of confounders from structural diversity in the prediction of personal social reputation, we conduct a series of matching experiments. 
After synthetically controlling possible confounding factors, we present compelling evidence that for individuals with an equal number of followers, same gender, and similar activity-related patterns, those whose followers come from more diverse social backgrounds (measured by structural diversity) are more likely to have higher social reputations.
Our work presents one of the first attempts to empirically study structural diversity in directed networks and demonstrates the potential utility of structural diversity in predicting personal social outcomes, and could also shed significant insights into the study of a range of social processes such as the diffusion of innovations, the spread of infectious diseases and the influence maximization problem.

\begin{figure}[!tb]
    \centering
    \includegraphics[scale=.75]{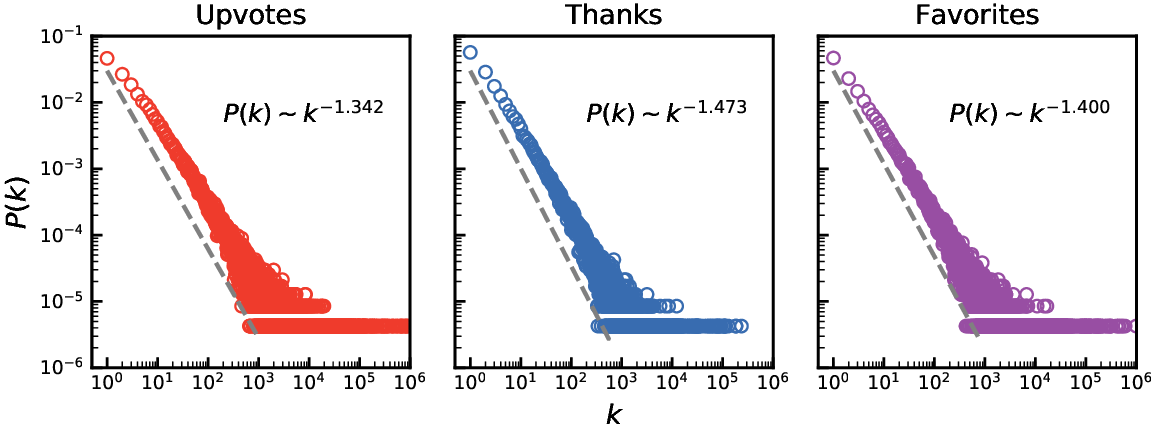}
    \caption{Distributions of the number of received upvotes, thanks, and favorites.
    For ease of visualization, quantities of zeros are not shown.
    The dashed grey line in each panel shows the power-law fitting for each distribution.}
    \label{fig:outcome_dist}
\end{figure}

\section{Results}

\subsection{Network data and social reputation index}

We collect data from Zhihu, a Chinese knowledge-sharing website which claims to have more than 200 million registered users.
On this platform, social ties are created when users choose to follow other accounts. Starting from a randomly selected user, we collect follower and followee lists of 234,834 users in a snowball sampling manner. 
These 234,834 users are hereafter referred to as \textit{ego} users since we know their complete followers and followees.
In addition, we also collect the followee lists of the ego users' followers (user accounts two hops away from the ego users).
The whole social network is then constructed based on the explicit social ties between user accounts. In total, the constructed network covers more than 10 million user accounts and 300 million directed social ties.
For these $\sim$ 230,000 ego users, we further collect their popularity data, including how many upvotes, thanks, and favorites they have received, which indicate their social reputations on the platform.
Other kinds of informative data are also collected, such as how many questions they have asked and answered, self-reported gender, followed topics, and questions.
Note that, all the collected data are based on public information on the platform and don't include any users with privacy restrictions.

The distributions of the number of received upvotes, thanks, and favorites are illustrated in Figure \ref{fig:outcome_dist}, where each point in the panel indicates the fraction or relative frequency $P(k)$ of users with a specific quantity equals $k$.
The dashed grey line in each panel shows the power-law fitting \cite{clauset2009power, alstott2014powerlaw} for each distribution, where the power-law exponents are 1.342, 1.473, and 1.400, respectively.
As the distributions span several orders of magnitude, the inequality of personal popularity is very striking. Considering the fact that these three popularity measures are highly correlated and very sparse, we adopt non-negative matrix factorization (NMF) \cite{lee1999learning, lee2000algorithms, cichocki2009fast, fevotte2011algorithms}–a widely used dimensionality reduction technique–to collapse them into a single measure which we term \textit{Social Reputation Index}
(see Methods section for details on the construction of social reputation index).

\begin{figure*}[!tb]
    \centering
    \includegraphics[scale=.6]{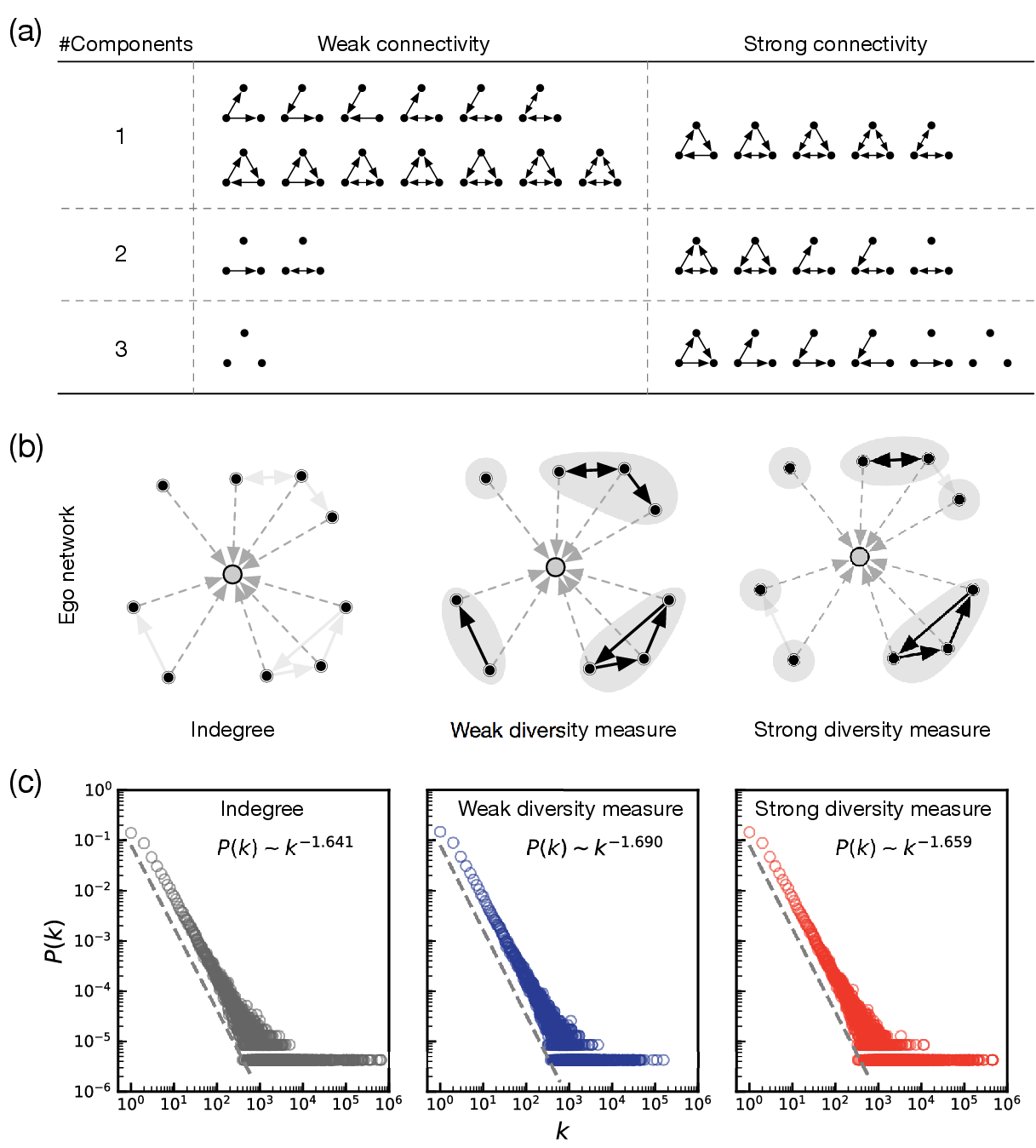}
    \caption{Weak and strong connectivity.
    (\textbf{a}) Projecting three-node directed networks into the number of weakly or strongly connected components depending on the connectivity patterns between nodes.
    %
    The first, second, and third rows
    correspond to situations where the number of weakly or strongly connected components is exactly one, two, and three, respectively.
    (\textbf{b}) Illustration of how the structural diversity measures are quantified compared with indegree.
    \textit{(left panel)} For the given ego network, the ego user and his/her followers are shown in grey and black dots, respectively, and the incoming links of the ego user are shown in dashed lines while the links among followers are faded in grey.
    \textit{(middle-right panels)} Illustration of 
    how weak and strong diversity measures are computed, with the informative links highlighted in black and weakly or strongly connected social components shown in shading areas.
    For the given example, the indegree of the ego user is 9 as he/she has 9 followers;
    weak diversity measure is 4 since there are only 4 weakly connected components formed by these 9 followers; 
    and strong diversity measure is 6 since there are 6 strongly connected components formed by these 9 followers.
    (\textbf{c}) Distributions of indegree, weak diversity measure, and strong diversity measure.
    For ease of visualization, quantities of zeros are not shown.
    The dashed grey line in each panel shows the power-law fitting for each distribution.}
    \label{fig:weak_strong_connectivity}
\end{figure*}

\begin{figure*}[!t]
    \centering
    \includegraphics[width=\hsize]{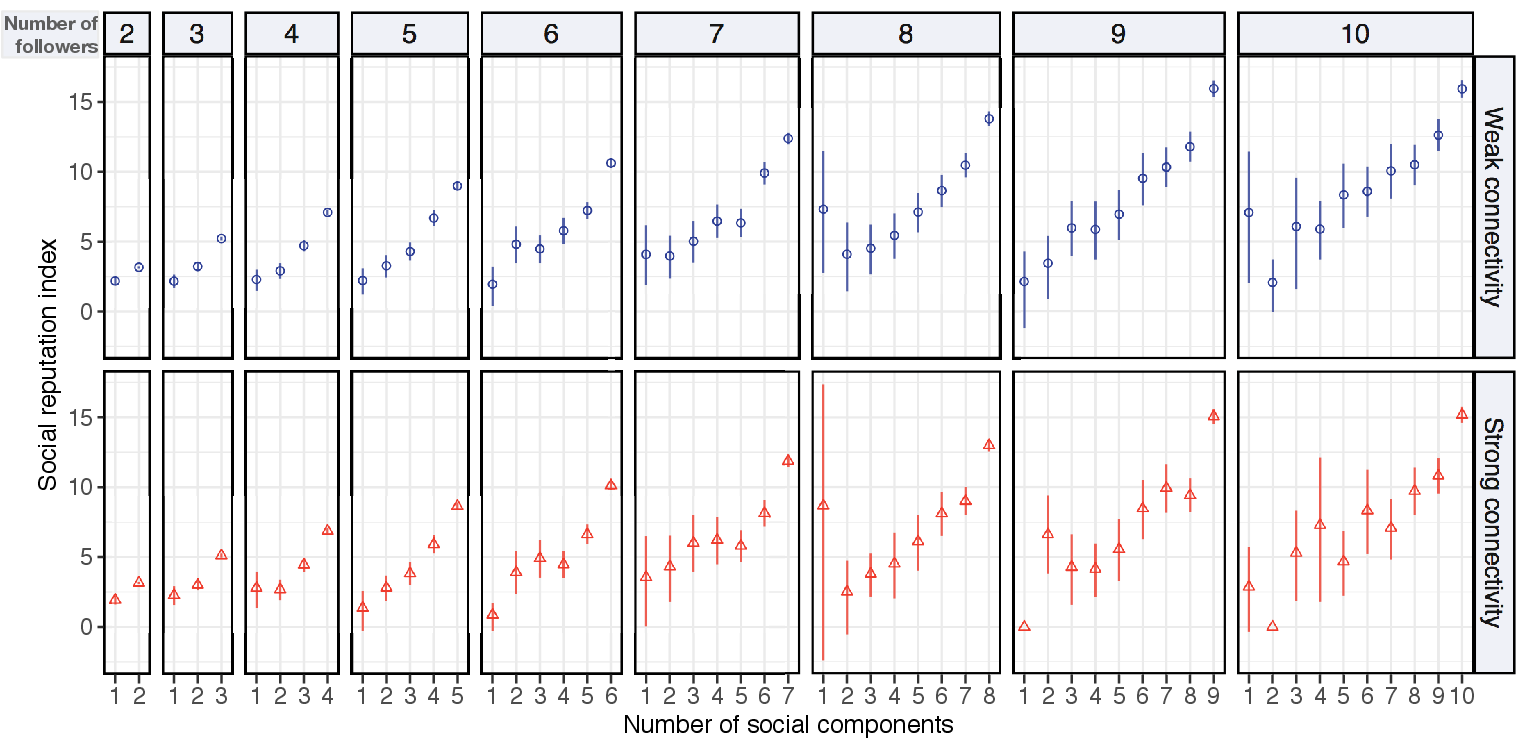}
    \caption{
    Social reputation index versus structural diversity.
    Social reputation index for two to ten-follower ego networks in terms of weak and strong diversity measures stratified by the number of followers.
    Error bars are 95\% confidence intervals.
    }
    \label{fig:weak_strong_diversity}
\end{figure*}

\subsection{Weak and strong connectivity}

To quantify the structural diversity of a given node in directed networks, we first consider the weak and strong connectivity between the neighboring nodes.
Note that for any two nodes \textit{u} and \textit{v} in a directed network, \textit{u} and \textit{v} are said to be weakly connected as long as there is a path linking \textit{u} and \textit{v} regardless of the direction of the path, and strongly connected if and only if there is at least a directed path from \textit{u} to \textit{v} as well as a directed path from \textit{v} to \textit{u}.
Therefore, according to the weak or strong connectivity patterns between nodes, a directed network can be decomposed into several social components.

There may exist multiple approaches to quantify the structural diversity of individuals, but the most simple and straightforward way would be the number of connected social components in one's contact neighborhood \cite{ugander2012structural, su2020experimental}.
Taking networks with three nodes as examples, 
Figure \ref{fig:weak_strong_connectivity}(a) shows how directed networks
are projected into different numbers of social components based on the weak or strong connectivity patterns.
For a constructed ego network with the ego node/user (we use the terms node and user interchangeably throughout the paper) located at the hub of the wheel,
Figure \ref{fig:weak_strong_connectivity}(b) further illustrates how the corresponding structural diversity measures are computed compared with indegree (number of followers).
For the given ego user in Figure \ref{fig:weak_strong_connectivity}(b), 
($i$) indegree is equal to the number of followers, which is 9 in the given example;
($ii$) weak diversity measure is equal to the number of weakly connected components in the contact neighborhood of the ego user, which is 4 in the given example;
($iii$) strong diversity measure is equal to the number of strongly connected components in the contact neighborhood, which is 6 in the given example. 
In fact, we have the property that weak diversity measure $\leq$ strong diversity measure $\leq$ indegree for any given ego user.
For these  $\sim$ 230,000 ego users, more than 91\% of them have less than 20 followers, and only 1,442 of them have more than 1,000 followers (Figure \ref{fig:weak_strong_connectivity}(c), left panel).
As shown in Figure \ref{fig:weak_strong_connectivity}(c), the distributions of their weak and strong diversity measures are also very highly skewed.
The dashed grey line in each panel shows the power-law fitting \cite{clauset2009power, alstott2014powerlaw} for each distribution, where the power-law exponents are 1.641, 1.690, and 1.659, respectively.
As expected, the social reputation index increases with all three metrics, with Spearman's rank correlation coefficients as 0.646 for indegree ($p<0.001$), 0.648 for weak diversity measure ($p<0.001$) and 0.647 for strong diversity measure ($p<0.001$).

\begin{figure}[!t]
    \centering
    \includegraphics[width=7cm]{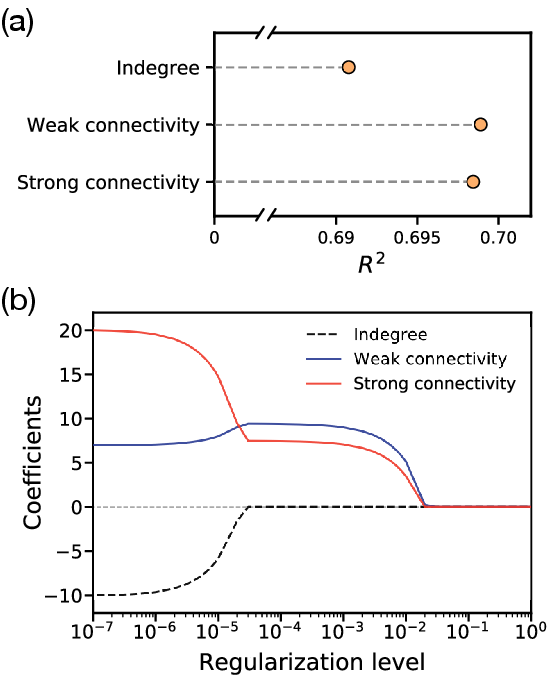}
    \caption{
    Prediction accuracy and regularization path.
    (\textbf{a}) Prediction accuracy ($R^2$) of social reputation by indegree and diversity measures (log-transformed) obtained via weak and strong connectivity.
     (\textbf{b}) Regularization path of the $L_1$-regularized linear regression model with indegree, weak and strong diversity measures (log-transformed) set as predictors.
    The log transformation is done by $\log_{10}(x+1)$ in each regression.
    }
    \label{fig:indegree_sd_comparison}
\end{figure}

With a closer look at individuals with an equal number of followers (indegree), we find that social reputation index generally grows monotonically with the increase of weak or strong diversity measure
(Figure \ref{fig:weak_strong_diversity}).
In other words, for individuals with the same number of followers, those whose followers come from more diverse social backgrounds (measured by weak or strong diversity measure) are likely to have higher social reputations.
Figure \ref{fig:indegree_sd_comparison}(a) summarises the prediction accuracy ($R^2$) of social reputation by indegree, weak diversity measure, and strong diversity measure through three separate ordinary least squares (OLS) regressions.
In each regression, the social reputation index is set as the dependent variable and each measure (log-transformed) is set as the sole independent variable.
As shown in the figure, structural diversity measures yield better predictions of online social reputation than indegree with $R^2$ values equal 0.691, 0.699, and 0.698 for indegree, weak diversity measure, and strong diversity measure, respectively.

To directly compare these measures, we further adopt $L_1$-regularized linear regression, which is also called least absolute shrinkage and selection operator (LASSO) regression.
LASSO regression is a standard model in sparse regression and has been widely used for simultaneous estimation and variable selection \cite{tibshirani1996regression, zou2006adaptive, zhao2006model, friedman2010regularization, yin2020measuring}
(see Methods section for more information on LASSO regression).
Note that, with the increase of the regularization level, LASSO would continuously shrink the coefficients of less important features to be zero \cite{zou2006adaptive, zhao2006model}.
Figure \ref{fig:indegree_sd_comparison}(b) shows the regularization path obtained from LASSO regression where indegree, weak and strong diversity measures (log-transformed) are set as predictors in the prediction of social reputation.
With the decrease of the regularization level, weak and strong diversity measures are first selected before indegree, which may imply that it's not how many followers one has but rather the structural diversity of one's followers that matters in capturing one's social reputation.

\begin{figure*}[!t]
    \centering
    \includegraphics[width=\hsize]{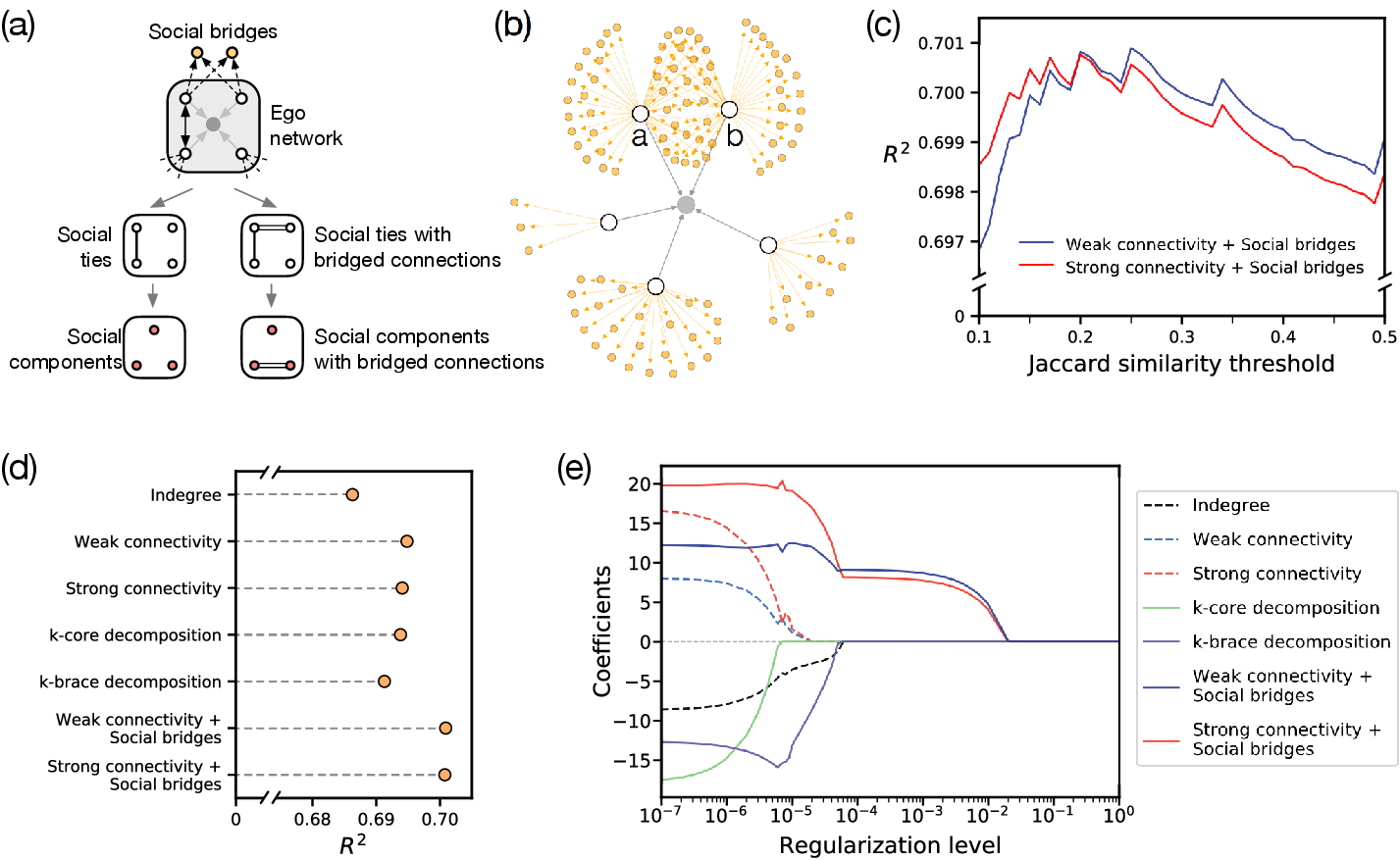}
    \caption{Social bridges.
    (\textbf{a}) Illustration of an ego user (grey dot) with four followers (black circles) and the projection from the original ego network to social components.
    Direct social ties between followers are highlighted in black, while social connections between followers and their followees are shown in dashed lines.
    In the given example, the top two followers have three followees (act as social bridges) in common: one is the ego user, and the other two are colored in orange; as such, a ``bridged connection" (double solid line) between the two followers is induced due to the function of social bridges. 
    Direct social ties lead a four-node connected neighborhood to three social components, whereas social bridges link two of the three social components. 
    In other words, the number of connected components becomes 2 after social bridges are considered.
    (\textbf{b}) An example ego network from the data with five isolated followers.
    In this example, social bridges provide enhanced ability to depict structural diversity as node a and node b share a large proportion of followees.
    (\textbf{c}) Prediction accuracy ($R^2$) of social reputation with the changing of Jaccard similarity threshold in the identification of bridged connections.
    (\textbf{d}) Prediction accuracy ($R^2$) of social reputation by diversity measures (log-transformed) obtained via different approaches.
    (\textbf{e}) Regularization path of the $L_1$-regularized linear regression model with several diversity measures (log-transformed) set as predictors.
    To facilitate the computation and make these measures comparable, we focus on ego users with no more than 30,000 followers.
    Diversity measures are log-transformed by $\log_{10}(x+1)$ in each regression.
    }
    \label{fig:social_bridges}
\end{figure*}

\subsection{Social bridges}

To quantify one's social context diversity from the view of component count, previous studies have focused exclusively on explicit and direct social ties in one's contact neighborhood.
However, as social neighbors provide a reliable way to infer personal characteristics \cite{mcpherson2001birds, al2012homophily, liben2007link, lu2011link, christakis2013social, liang2018birds}, it's reasonable to speculate that two individuals who are exposed to a similar set of users will tend to be similar with each other, even without a direct social tie between them.
Therefore, the common followees between them could act as \textit{social bridges} that implicitly ``link'' two unconnected individuals or social components in the network \cite{dong2018social}.

Figure \ref{fig:social_bridges}(a) illustrates the process of how we apply social bridges to capture the implicit structural diversity of ego nodes.
Figure \ref{fig:social_bridges}(b) gives another example with all the followers (denoted by black circles) isolated (i.e, no direct social ties between the followers). However, after social bridges are taken into consideration, user a and user b are likely to form one connected component
(see Figures \textcolor{blue}{S3} and \textcolor{blue}{S4} of the Supplementary Materials for more examples).
Specifically, for two followers $i$ and $j$ of an ego user, we adopt Jaccard similarity of their followee sets to determine whether there is a ``bridged connection" between them: $JaccardSim(i,j) = |F_i \cap F_j|/|F_i \cup F_j|$, where $F_i$ and $F_j$ denote the followee sets of $i$ and $j$, $\cap$ and $\cup$ denote the intersection and union operators of two sets, and $|\bullet|$ denotes the size of a given set.
A bridged connection exists between $i$ and $j$ when $JaccardSim(i,j)$ is larger than a given threshold. 
A small threshold indicates that a bridged connection exists between two individuals as long as they share a small fraction of followees, while a large threshold requires two individuals sharing a large proportion of followees to determine the existence of a bridged connection.

Note that the bridged connection meets the requirement of strong connectivity between two nodes.
Therefore, if two nodes share a bridged connection, they are also weakly and strongly connected.
After social bridges are considered, the number of weakly and strongly connected components are termed the enhanced diversity measures via social bridges, denoted as `Weak connectivity+Social bridges' and `Strong connectivity+Social bridges', respectively.
For ease of computation, ego users with more than 30,000 followers (i.e., 88 out of 234,834 ego users) are omitted in the analysis.
Figure \ref{fig:social_bridges}(c) shows the prediction accuracy ($R^2$) of social reputation by diversity measures via social bridges with the change of Jaccard similarity threshold.
Note that each $R^2$ value is obtained via OLS regression with the social reputation index set as the dependent variable and diversity measure (log-transformed) under each condition set as the sole independent variable.
For weak connectivity, a threshold around 0.25 achieves the best performance; for strong connectivity, a threshold around 0.2 achieves the best performance.

In Figure \ref{fig:social_bridges}(d), we summarise the best prediction accuracy ($R^2$) achieved by each diversity measure via different approaches, including two approaches that are proposed for undirected networks--$k$-core decomposition and $k$-brace decomposition \cite{ugander2012structural}.
Similar to above, to facilitate the computation and make these diversity measures comparable, only ego users with no more than 30,000 followers are included in the analysis.
We also make some slight changes to make $k$-core and $k$-brace decomposition applicable in directed networks (see Methods section for details).
Specifically, $k$-core and $k$-brace decomposition achieve the best prediction accuracy ($R^2$) when $k$ is set to be 2 and 1, respectively.
As we can see from Figure \ref{fig:social_bridges}(d),
social bridges provide the most additional ingredient to predict personal online social reputation, and diversity measure via weak connectivity and social bridges yields the best prediction performance.
Furthermore, Figure \ref{fig:social_bridges}(e) shows the regularization path of the $L_1$-regularized linear regression (LASSO regression) model with several diversity measures (log-transformed) set as predictors.
%
%
As shown in the figure, with the decrease of the regularization level, diversity measures via social bridges are selected before other diversity measures, which further suggests the effectiveness of social bridges in the prediction of online social reputation.

\subsection{Robustness analysis}

\begin{figure*}[!t]
    \centering
    \includegraphics[width=\hsize]{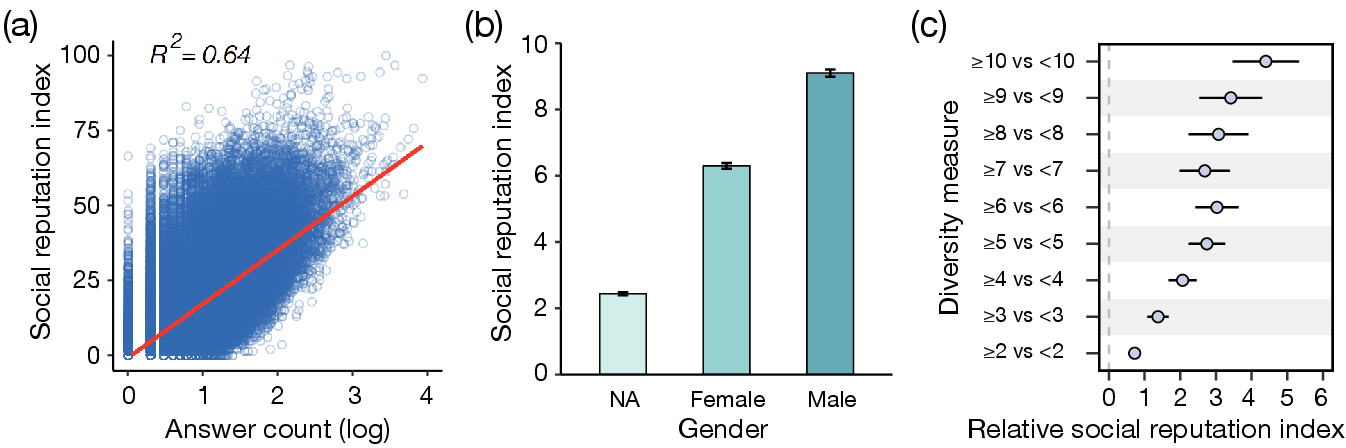}
    \caption{
    Robustness analysis.
     (\textbf{a}) Scatter plot of social reputation index versus answer count, where red line indicates the linear fit. 
    (\textbf{b}) Gender differences in terms of social reputation index, where NA refers to users whose gender is not disclosed.
    (\textbf{c}) Matching results where the treatment assignment is set as whether the quantified structural diversity measure is larger than or equal to $m$ ($m \in [2, 10]$).
    Error bars are 95\% confidence intervals.
    }
    \label{fig:robustness}
\end{figure*}

In the previous sections, we have shown that individuals with higher levels of structural diversity tend to have higher online social reputations.
However, the positive correlation between structural diversity and online social reputation may be biased by other factors.
For example, as shown in Figure \ref{fig:robustness}(a), answer count is also positively correlated with online social reputation (\textit{Spearman's} $r=0.858$, $p<0.001$).
In other words, the more answers a user has contributed to the knowledge-sharing community, the higher social reputations he/she tends to have on the platform.
Therefore, personal online social reputation may also be induced by contributed answers rather than merely the structural diversity of individuals.
We also find that gender could be another potentially confounding factor, as male users tend to receive higher social reputations than female users and users of unknown gender on the platform (\textit{one-way ANOVA}, $F(2, 234831)=6636.8$, $p<0.001$) (Figure \ref{fig:robustness}(b)).
Combining with several other activity-related factors, such as question count and article count, a series of matching experiments are then conducted to distinguish the effects of structural diversity from these possible confounding factors in capturing personal social reputation.

Propensity score matching (PSM) is a widely used method for matching experiments in the literature \cite{rubin2001using, stuart2010matching, aral2009distinguishing, aiello2017beautiful, kumar2018community, way2019productivity, ho2011matchit}. 
The key intuition of PSM is to match the treatment group with a control group whose members don't receive the treatment but are statistically indistinguishable or at least only marginally different (within a reasonable limit) from the treatment group on all observable covariates.
We use the quantified diversity measure via weak connectivity and social bridges to depict the structural diversity of users in the network.
Here in our setting, a user is said to be treated (treatment group) if his/her diversity measure is larger than or equal to $m$ ($m$ is a given integer), otherwise untreated (control group).
We do exact matching on indegree and gender (i.e., matched pairs have equal indegree and the same gender) and propensity score matching on other covariates (see Table 
\textcolor{blue}{S6}
of the Supplementary Materials for detailed covariates controlled in the matching experiments).
The difference of social reputation index in each matched pair (treatment - control) indicates the relative social reputation induced by the increase of structural diversity.

After matching, we obtain a well-balanced dataset with all standardized mean differences between the treated and untreated groups being less than 0.25.
Figure \ref{fig:robustness}(c) shows the differences of social reputation index between matched pairs (for space constraint and simplicity, we only present results when $m$ is set to be in the range [2, 10]).
As shown in the figure, after ruling out several potentially confounding factors, users with higher levels of structural diversity (assigned as treated users in the matching experiments) would still tend to have higher online social reputations (\textit{paired t-test}, $p<0.001$ for all matching experiments).
Taken together, after rigorously controlling possible confounders, matching experiments provide compelling evidence for the role of structural diversity in predicting personal online social reputation.

\section{Discussion}

The advent of social networking sites and knowledge-sharing platforms has radically shifted the way we consume information, acquire knowledge, and exchange ideas. 
As social ties among individuals provide the primary pathways along which interactions occur, the way we are connected and embedded in social networks is thought to affect various personal social outcomes, ranging from personal health to socio-economic characteristics.

Taking advantage of network data which cover more than 10 million users (including more than 230,000 ego users) from an online knowledge-sharing platform, our study highlights the importance of structural diversity in online social networks and suggests an alternate perspective for people to accumulate their social capital, for policy makers to make appropriate interventions and for market operators to set up effective campaigns.
Our findings are also of prime significance to understand why network structure matters in a range of social and economic domains.
For example, individuals who are located in a diversified social context are generally accessible to novel information and ideas, which has important implications for viral marketing and fake news research.
Moreover, as we live in such a connected world of overloaded information, how do we aggregate opinions around us (e.g., adopt information from diverse backgrounds) to arrive at reliable and accurate estimation is not only crucial to make better decisions but also important to improve the collective intelligence.

Our work is subject to a number of limitations.
Our results are the product of one study based on the collected data from an online knowledge-sharing platform. Therefore, additional studies are needed to validate our findings in other kinds of social applications or domains.
In this paper, we use snowball sampling during the data collection, but this approach may over- or under-sample some data and induce bias to the results.
To quantify the structural diversity of individuals, this work mainly considers the number of connected social components in one's social realm, but other network factors, such as the weight of ties and other structural properties, may help to achieve that goal and are also worth exploring in future works.
Although social bridges help to capture the implicit structural diversity of individuals, the computation will require a large amount of computing time and memory resources, especially for users with a large number of followers.
In our study, the matching experiments have already accounted for several observed characteristics of users, but due to the limitation of data availability, the estimation results may still be biased without properly controlling those unobserved or unmeasured confounding factors. 
We emphasize that our results are built upon correlation analysis on observational data, thus not implying causality sufficiently.
The current study mainly considers social ties induced by following relationships, but social connections induced by behavioral changes, such as comment or retweet, may provide another way to infer the interrelationships between individuals and thus are worth exploring in future studies.

\section{Methods}

\subsection{Construction of social reputation index}

The current study adopts non-negative matrix factorization (NMF) to construct the social reputation index from the original data.
NMF is an unsupervised approach for dimensionality reduction with the constraints that the input and output matrices don't contain any negative elements \cite{lee1999learning, lee2000algorithms, cichocki2009fast, fevotte2011algorithms}. 
Specifically, for a given non-negative data matrix $V$, NMF attempts to find an approximate factorization:
$V \approx W \cdot H$
such that the original matrix $V$ can be decomposed into two non-negative submatrices $W$ and $H$, with the goal of minimizing the reconstruction error between $V$ and $W \cdot H$.

In our setting, the decomposition of an original matrix with $n$ users and $m$ dimensions of features can be decoded as $V_{n \times m} \approx W_{n \times r} \cdot H_{r \times m}$, where $r$ is a given parameter prior to the matrix decomposition and indicates the expected dimension after NMF.
Here in our study,
$V_{n \times m}$ is the original data matrix with $n$ equals  $234,834$ (i.e., the number of ego users in the sample) and $m$ equals 3 as there are three types of popularity measures (i.e., number of received upvotes, thanks, and favorites),
whereas $r$ is set to be 1 as we aim to obtain a single measure of social reputation. 
After NMF, the reduced submatrix $W_{n \times r}$ is denoted as the social reputation index.

In practice, these three popularity measures are first log-transformed by $\log_{10}(x+1)$ and then fed into NMF simultaneously. 
Once the social reputation index (the first submatrix $W$ after NMF decomposition) is obtained, we normalize it to the range [0, 100] as follows: 
given a vector $W$ indicating the social reputation index, every element of $W$ is normalized as 
$W_i= 100 \times {\big(W_i-W_{min}\big)}/{\big(W_{max}-W_{min}\big)}$,
where $W_{max}$ and $W_{min}$ are maximum and minimum values of $W$, respectively.

\subsection{LASSO regression}

Generally speaking, OLS regression yields nonzero estimates to the coefficients of all the features.
But LASSO adds the $L_1$ penalty of the coefficients (i.e., the sum of the absolute value of the estimated coefficients) as the regularization term to the loss function of OLS regression (see Refs. \cite{zou2006adaptive, zhao2006model, friedman2010regularization} for further technical details).
Therefore, when the regularization level is set to be zero, LASSO regression is equivalent to OLS regression.
With the increase of the regularization level, LASSO would gradually force the coefficients of less important features to be zero.
When the regularization level becomes sufficiently large, all the estimated coefficients will be zero.

\subsection{k-core and k-brace decomposition}

$k$-core and $k$-brace decomposition have been shown useful in depicting structural diversity in undirected networks \cite{ugander2012structural}.
Considering the fact that users with completely isolated followers tend to have higher social reputations than their counterparts, node removal in $k$-core and $k$-brace decomposition would not work here.
In this regard, we make some slight changes to make $k$-core and $k$-brace decomposition applicable in current scenario.
In the contact neighborhood formed by the followers of an ego user, only edges that are affiliated to specific nodes are removed (instead of node removal).
For $k$-core decomposition, we use the degree (i.e., indegree+outdegree) of nodes to do the decomposition process:
edges that are affiliated with nodes whose degrees are less than $k$ are repeatedly removed.
For $k$-brace decomposition, we transform the directed networks to undirected ones and then implement the decomposition process:
edges whose two endpoints share less than $k$ neighbors are repeatedly removed.
After decomposition, the number of (weakly) connected components in the decomposed contact neighborhood is named the diversity measure via $k$-core or $k$-brace decomposition.

\section*{Supplementary Materials}

The Supplementary Materials are available at:

\noindent
\url{https://downloads.spj.sciencemag.org/research/2021/9831621.f1.pdf}.

\printbibliography


\end{document}